# Probability Risk Identification Based Intrusion Detection System for SCADA Systems

Thomas Marsden, Nour Moustafa, Elena Sitnikova, and Gideon Creech

School of Engineering and Information Technology, UNSW Canberra, Australia
thomas.marsden@defence.gov.au,nour.moustafa@unsw.edu.au,e.sitnikova@adfa.edu.au,g.creech@adfa.edu.au

**Abstract.** As Supervisory Control and Data Acquisition (SCADA) systems control several critical infrastructures, they have connected to the internet. Consequently, SCADA systems face different sophisticated types of cyber adversaries. This paper suggests a Probability Risk Identification based Intrusion Detection System (PRI-IDS) technique based on analysing network traffic of Modbus TCP/IP for identifying replay attacks. It is acknowledged that Modbus TCP is usually vulnerable due to its unauthenticated and unencrypted nature. Our technique is evaluated using a simulation environment by configuring a testbed, which is a custom SCADA network that is cheap, accurate and scalable. The testbed is exploited when testing the IDS by sending individual packets from an attacker located on the same LAN as the Modbus master and slave. The experimental results demonstrated that the proposed technique can effectively and efficiently recognise replay attacks.

## 1 Introduction

Research into the security of SCADA systems has grown in recent years, as the potential damage to critical infrastructure including gas, electricity, water, traffic and railway, and/or loss of life and subsequent risk to state security have been realised [19]. SCADA refers to a system of computers and programmable logic controllers (PLCs) that control and monitor industrial plants, processes and machinery [19, 11]. It enables technicians and engineers to supervise and take control of systems remotely [25]. SCADA is commonly employed in systems that are considered critical infrastructure, essential services, in which society re- lies on in day to day life. More specifically, critical infrastructure is defined by TISN as "Those physical facilities, supply chains, information technologies and communication networks, which if destroyed, degraded or rendered unavailable for an extended period, would significantly impact on the social or economic well-being of the nation, or affect Australia's ability to conduct national defence and ensure national security" [7, 27]. Unfortunately, most studies have unveiled that security is an afterthought at best in SCADA systems [19]. Various steps taken to mediate the weaknesses have been suggested and are discussed in section 2. We suggest the implementation of an IDS [24, 23]. Commonplace in traditional information technology networks, IDSs are deployed to alert a systems administrator when malicious activity is detected on their network [10]. However, their adoption in SCADA networks has been limited due to a lack of security best practice [15].

Replay attacks are network based attacks where valid data is repeated at a target to cause malicious effect. They are executed by a MitM or intended source, and due to the unencrypted and unauthenticated norm of Modbus TCP, replay attacks are highly effective in exploiting SCADA as these attacks attempt to disrupt the flow of traffic between source and destination [19]. The unencrypted nature permits a replay attack to modify the contents of a Modbus TCP packet [12]. This would manifest itself as modifying values stored in registers, for example if a traffic authority were to take manual control of a set of traffic lights in an accident, a packet directing red lights to 'turn on' could be changed to say 'turn off' or never arrive at all [11]. The unauthenticated nature of Modbus TCP means that any user on the same LAN as a Modbus slave is able to access memory values of the PLCs and write values to create similar effects [11].

In this study, we suggest a Probability Risk Identification-based (PRI-IDS) for the Modbus TCP protocol named BusNIDS, with the aim of detecting replay attacks. It will achieve this through characterising data with pre-determined risk values, caching periods of data and generating risk values for those cached periods. Caches with risk levels outside of one standard deviation as a threshold from the mean are flagged as potential replay attacks. All information collected from packets is available as header information from the Application Data Unit (ADU) and Protocol Data Unit (PDU), this means that we can deploy this technique in realtime on an encrypted network [9]. A small scale industrial control network detailed in section 4, used for running attacks and testing the IDS and obtaining results comparing with three peer IDS techniques.

The key contributions in this paper are summarised as follows:

- We develop a Modbus TCP IDS that can conduct packet analysis and detect replay attacks.
- We develop a SCADA testbed which is low-cost and simple to configure for evaluating the proposed IDS.
- We compare the performance of the proposed technique with three existing techniques, showing its superiority.

## 2 Background and Literature Review

This section explains the background and previous studies related to the Modbus TCP protocol, IDS for SCADA systems, establishing testbeds for the purpose of IDS research and provides a contextualised review based on aims of the research.

SCADA systems feature a number of security challenges. Bartman *et al.* [5] describe SCADA networks with nine primary threat vectors. They are replay attacks, MitM, brute force, dictionary, eavesdropping, Denial of Service (DoS), war dialling, default credentials and data modification. They establish that implementing IPsec to transport data, AES encryption natively implemented into SCADA protocols and the use of IDSs reporting to a centralised Security Information and Event Management (SIEM) are three of the most effective methods to secure a SCADA network [5]. We discuss that the implementation of an IDS that analyses Modbus TCP header data to detect risks, and as such can improve the security posture of a system by working in tandem with encryption [9].

A review of current literature regarding IDS solutions that support Modbus TCP is required to form a baseline knowledge and assists in revealing where novelties for IDS research lay. Yüksel *et al.* [30] develop an anomaly based engine that they test utilising Modbus TCP, Modbus RTU and Siemens S7 datasets. The anomaly based engine is analysing individual TCP packets, determining a probability that the packet is normal traffic based on features of the packet (Source IP, Destination IP, Functions Codes, Read/Write values, etc). Packets with probability deviation beyond a set limit are categorised as threats and generate an alert. The engine is trained faster than existing datasets and is able to detect malicious traffic faster by a factor of 30% over its closest competitor. The trade off for these results is that when using live data rather than a pre- recorded dataset, each packet takes 0.7ms to parse [30]. Additionally, deploying the technique is limited in scope, as payload data is potentially encrypted and data available to the engine is limited. We demonstrate that through the inclu- sion of packet analysis and Modbus TCP session analysis, that a Modbus TCP IDS provides an improvement to the detection of malformed packets and replay attacks over existing algorithms trained using machine learning datasets.

Research in the SCADA field has been invested into developing accurate testbed systems at reduced cost [2]. Ahmed *et al.* [1] developed a SCADA system testbed for cybersecurity and forensic research which models a gas pipeline, wastewater treatment plant and power distribution centre. Each plant is modelled using a different brand of in-service PLC, and thus a separate protocol from each plant to a centralised HMI. Due to the testbed utilising currently employed protocols and hardware, vulnerabilities discovered in it are likely to be reproduced in deployed systems [1]. An alternative approach is suggested by Morris *et al.* [2]. The research showed that a completely simulated and virtualised system could compare in accuracy of control and reception of data to a physical system. Genge *et al.* [13] propose a hybrid simulation and emulation framework. PLCs and HMIs are emulated on the Emulab platform, allowing for virtual remapping of physical hardware. The distinct advantage of this method is that it allows for rapid expansion of a system, i.e., 100s of PLCs can be stood up rapidly in the system [13]. The authors showed that a physical test network has the attraction of being cheap, accessible and accurate. However, it does not demonstrate the same scalability as virtualised alternatives.

There are multiple pieces of research dedicated to datasets for the testing of IDSs. Morris *et al.* [22] have created datasets for a laboratory gas pipeline and water storage tank. Lemay *et al.* [18] take the view that the current datasets are too limited and are slanted towards the detection of malformed packets. Thus, they devise a set of network data that includes normal operation, operator manipulation, Metasploit exploits, unauthorised remote read/write commands, network scanning and exploitation of their covert Modbus TCP channel for command and control (C2) [18]. They simulated a small electrical network as a plant. The datasets provided by Lemay *et al.* [18] are publicly available and well rounded. The authors tested the Modbus TCP IDS with individual malicious commands interceding normal network operation.

## 3 Probability Risk Identification (PRI-IDS) Technique

The PRI-IDS technique is designed based on computing a risk level for network data. The following process is summarised as a flowchart in Figure 1 that demonstrates how our technique works for detecting abnormal events from SCADA data.

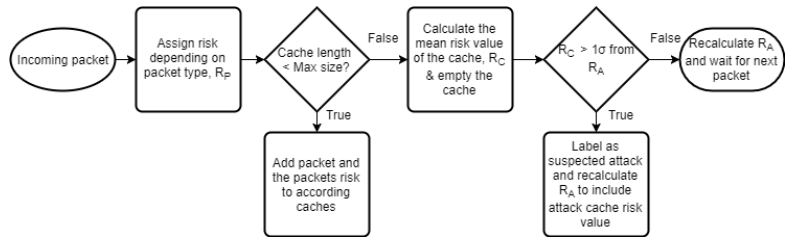

**Fig. 1.** Flowchart summarising the PRI-IDS technique

More importantly, we assign a base risk probability, $R_P$ (Risk value of a packet), to each packet depending on its function code according to Table 1. The risk values assigned to each function code are dependant on the potential effect that the function code has on the system and can be changed depending on the environment that the IDS is deployed in. Each risk value is assigned in real-time as packets are processed. The $R_P$ value is increased an additional amount if the packet is identified as erroneous. The magnitude of risk values are dependent on the system in which the IDS is configured. As there are few cases in which the traffic light system should see write requests, these packets are more heavily weighted.

A configurable number of packets (CACHE_MAX_SIZE) are appended to a cache. The average risk probability value for the cache is calculated as $R_C$ (Risk value of a cache). Once the cache reaches CACHE MAX SIZE, the risk value for that cache is stored and used to calculate a moving average, $R_A$ (Moving average risk). The cache is cleared and progressively filled to its maximum size.

At every cache update, the $R_C$ values are checked. Values which fall greater than $1\sigma$ (Standard deviation, which is a threshold of identifying malicious events) from the mean $P_{RC}$ are flagged as potential replay attacks. The corresponding packets which were cause for the deviation are also flagged and made available to the operator. Caches which cause flags to be raised are stored to file in .PCAP formats for further analysis and added as a stored cache. On cache updates, the current cache is also checked against stored caches for matches in packet risk data.

**Table 1.** Risk Allocation for Modbus Function Codes

| Function type | | | Function name | Function code | Risk |
|---|---|---|---|---|---|
| Data Access | Bit access | Physical Discrete Inputs | Read Discrete Inputs | 2 | 0.1 |
| | | Internal Bits or Physical Coils | Read Coils | 1 | 0.5 |
| | | | Write Single Coil | 5 | 0.9 |
| | | | Write Multiple Coils | 15 | 0.9 |
| | 16-bit access | Physical Input Registers | Read Input Registers | 4 | 0.1 |
| | | Internal Registers or Physical Output Registers | Read Multiple Holding Registers | 3 | 0.5 |
| | | | Write Single Holding Register | 6 | 0.9 |
| | | | Write Multiple Holding Registers | 16 | 0.9 |
| | | | Read/Write Multiple Registers | 23 | 0.9 |
| | | | Mask Write Register | 22 | 0.5 |
| | | | Read FIFO Queue | 24 | 0.1 |
| | File Record Access | | Read File Record | 20 | 0.1 |
| | | | Write File Record | 21 | 0.5 |

## 4 Testbed and implementing proposed PRI-IDS technique

The testbed developed to run as a Modbus TCP network has three primary components. These components are the Raspberry Pi (RPi) running the openPLC software, traffic light program running on the PLC with a physical traffic light circuit interfacing via the RPi General Purpose Input Output (GPIO) pins, and a Windows 10 laptop running as a SCADA master via ScadaBR.

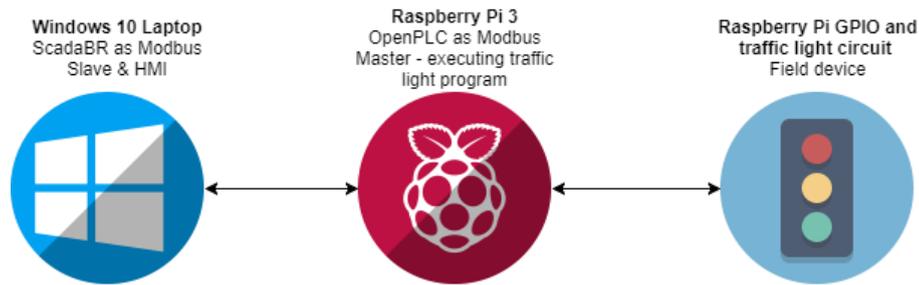

**Fig. 2.** Diagram of components in the SCADA testbed. Assembled using icon sets available at [16]

The openPLC software runs on multiple platforms as an effort to build a standardised open source PLC solution [3]. The software can be compiled on a RPi and achieves similar robustness to that of a commercial PLC [21]. At default, openPLC on the RPi operates with a 50ms scan rate, suitable for gas, water and traffic control SCADA applications [21]. Programs can be uploaded to the PLC through a local web interface. In addition to openPLC, the RPi runs the current iteration of the IDS. The openPLC was programmed with ladder logic to create a simple timer based traffic light program [8].

Fixed memory addresses in the PLC can be accessed via the GPIO pins on the RPi. A traffic light program and accompanying circuit have been constructed as a technique to visualise the effect of replay attacks on the PLC. For example, an attacker can intercept a valid write coils command and replay it at a later time, changing the intended state of the traffic lights [20]. In the physical circuit and HMI, this would present itself as changing the currently active light or disabling them completely. Figure 3 details the physical construction of the field device.

ScadaBR was used as a Modbus master and HMI, accessible via a web interface hosted by the Apache web-server. The laptop used to operate ScadaBR ran Windows 10 with an Intel Core i5-6200U at 2.3GHz and 8192MB of DDR4 RAM. ScadaBR is a fully-fledged SCADA master that supports multiple protocols, PLCs, sensors and custom HMIs [29].

The IDS is built in Python 2.7. There are a number of advantages to developing with Python 2.7. The interpreted nature of Python 2.7 enables live testing of code. Commands can be run individually in the interpreter before implementing them into a full script [28]. The Scapy framework is natively built for Python 2.7 and enables full network stack packet reading, writing and sniffing.

The Modbus TCP extension for Scapy enables the author to view packets in terms of Modbus commands rather than hexadecimal data dumps. It is a pivotal tool in the development of the IDS and for testing. It will be used for packet forging to test the IDS [6]. The IDS operates locally on the PLC due to Linux support for Scapy framework being more stable than alternatives. It reports when malformed Modbus TCP packets are sent to the PLC.

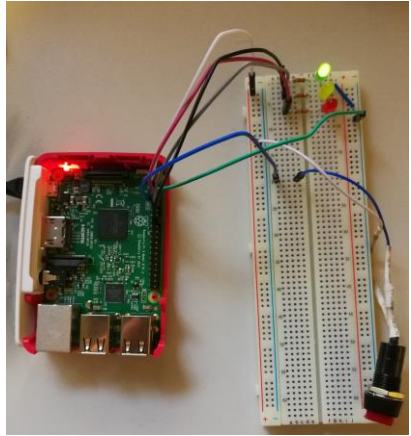

**Fig. 3.** Traffic light circuit connected to RPi PLC via GPIO

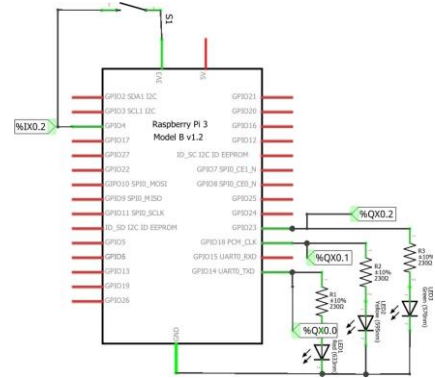

**Fig. 4.** Circuit diagram of traffic light and RPi PLC connections

## 5 Experimentation and Discussion

### 5.1 Dataset used for experimentation

A simulated attacker is used to test the IDS through a collection of Scapy commands that transmit Modbus TCP packets. These packets are considered the dataset that is used to evaluate the IDS and are sent directly to the PLC. Packets in the dataset are classified as either normal or attack packets. To generate the dataset commands, standard operational traffic between the SCADA master and slave was recorded. The operational traffic featured reading addresses from the PLC, and a write to the PLC to simulate manual override of the system. The traffic was converted to equivalent Scapy commands for repeatability and attacks were added. The malicious commands simulate replay attacks and change the state of coils that are actively used in the system in an unauthorised manner.

### 5.2 Evaluation metrics and testbed description

The IDS is characterised by the metrics accuracy, Detection Rate (DR) and False Positive Rate (FPR). Each of the metrics rely on the state of packets passed through the IDS, these states are defined as the terms true positive (TP), true negative (TN), false positive (FP) and false negative (FN). A packet is classified as TP if it is an attack packet that is correctly identified and thus, TP is the total number of TP classified packets. TN is the total number of packets that the IDS correctly classifies as normal traffic. FP refers to the number of packets that are normal however were classified as attacks by the IDS. FN is the number of attack packets that the IDS incorrectly classifies as normal traffic [24, 23]. The IDS metrics are calculated as functions of these terms according to equations 1,

2 and 3 below.

$$Accuracy = \frac{TP + TN}{TP + TN + FP + FN} \quad (1)$$

$$DetectionRate = \frac{TP}{TP + FN} \quad (2)$$

$$FalsePositiveRate = \frac{FP}{FP + TN} \quad (3)$$

In addition to testing the performance of the IDS, the performance of the Modbus TCP testbed is also considered. The testbed network was operated for a period of 1 hour. Every 10 minutes, write requests to the PLC were made and values in memory were altered. When write requests weren't being made, the PLC was being continuously polled with read requests.

**Table 2.** Errors in Modbus TCP traffic over the testbed for 88125 polls.

| Polls | OK    | Errors | Error % |
|-------|-------|--------|---------|
| 88125 | 88119 | 5      | 0.0057  |

Over the hour, the testbed provided a high level of accuracy. Table 2 displays the errors in Modbus TCP traffic on the testbed over an hour time period.

### 5.3 Algorithm performance compared with three techniques

The performance of proposed algorithm technique was assessed in terms of the overall detection rate and false positive rate, listed in Table 3 and shown in Figure 5. It is clear that our algorithm achieves the highest DR with roundly 83.7% and the lowest FPR with about 15.3% compared with the techniques [14] of K-Nearest Neighbour, Naïve Bayes and Random Forest . Performance of the alternative techniques used results generated by Hassan *et al* with technique parameters tuned according to their work.[14]

**Table 3.** Comparison of IDS algorithm performances

| Technique | DR | FPR |
|---|---|---|
| K-Nearest Neighbour (KNN) [14] | 55.3% | 44.8% |
| Naïve Bayes (NB) [14] | 44.4% | 52.6% |
| Random Forests (RF) [14] | 60.5% | 38.4% |
| Probability Risk Identification (PRI) | 83.7% | 15.3% |

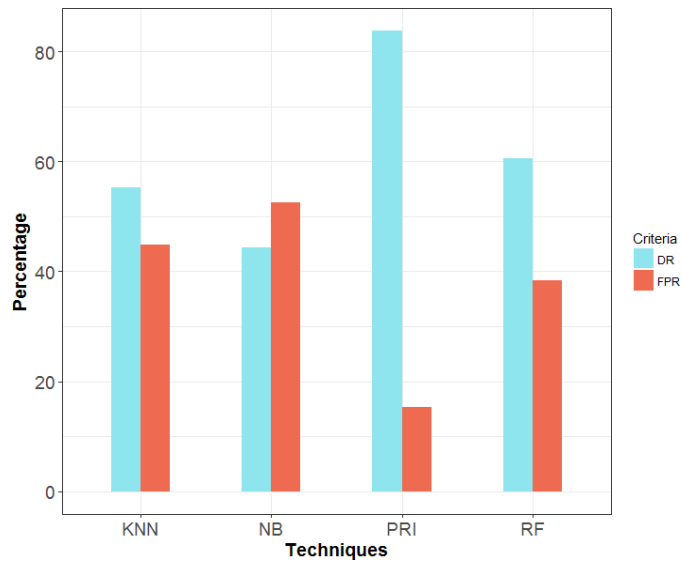

**Fig. 5.** Comparing performances of four IDS techniques

The evaluation criteria for machine learning algorithms in IDSs are DR and FPR. Clearly, the results for DR and FPR are improvements over the competing techniques shown in Table 3 and discussed in Hassan's work. The PRI-IDS produces improved results over the competing algorithms for Modbus TCP due to the customisable risk assignment to individual packets, depending on the potential process of each technique being used for testing. In more details, the K-Nearest Neighbour technique applies a majority vote function between neighbour data points, but as there is relatively a similarity between normal and abnormal data, it cannot achieve better than the PRI-IDS while running in realistic testbed environment [26]. Similarly, the Naïve Bayes and Random Forest techniques cannot find a clear difference between abnormal and normal observation of SCADA data [4, 17].

Ultimately, The PRI-IDS computes the probability of network observations with an accumulating likelihood for prior information, and this finds clear deviations between legitimate and suspicious observations compared with the three techniques. It means that clear distinction can be made between what is desirable and undesirable traffic at the packet level. However, the detection algorithm is susceptible to long, sustained attacks where an attacker transmits packets at a rate that builds the moving average up to a high risk level, allowing high risk packets to be sent without triggering a detection.

## 6 Conclusion

We discussed a Modbus TCP PRI-IDS capable of detecting replay attacks passively at a 28.4% better rate than the next best algorithm, Random Forest, and a corresponding testbed that is cheap, accurate and scalable. Ultimately, we developed an IDS technique designed to detect replay attacks that is more desirable to implement within a Modbus TCP network than alternatives. The PRI-IDS is written in Python 2.7 and relies on the Scapy network packet manipulation framework. The project is relevant due to the growth in Modbus TCP for communcations within SCADA networks. Additionally, the fragility of many SCADA systems means that implementing a passive solution into a network is often preferred over an active or in-line solution. In future work, we will implement the PRI technique into a machine learning environment to test it directly with existing SCADA IDS datasets to determine its performance. It will provide a known and widely used benchmark to test against and will enable fine-tuning of the technique. Once the PRI technique is fine-tuned, it will be reimplemented into a real-time environment to show the increase in performance against the initial iteration of the technique in an IDS.